\documentstyle[preprint,aps]{revtex}

\newcommand{\partialslash}{/\hspace{-2mm}{\partial}}
\newcommand{\Aslash}{/\hspace{-2.7mm}A}
\newcommand{\bslash}{/\hspace{-2mm}b}
\newcommand{\Pislash}{/\hspace{-2.7mm}\Pi}
\newcommand{\im}{\mbox{Im}}
\newcommand{\re}{\mbox{Re}}
\newcommand{\Tr}{\mbox{Tr}}

\begin{document}

\draft

\preprint{
\begin{tabular}{r}
HIP-2000-52/TH\\
FISIST/15-2000/CFIF\\
Guelph Mathematical Series 2000-301\\ \\
\end{tabular} }

\tighten

\title{Radiatively Induced Lorentz and $CPT$ Violation in Schwinger
 Constant Field Approximation}

\author{M.
Chaichian$^1$\renewcommand{\thefootnote}{\dagger}\footnote{E-mail:
chaichia@rock.helsinki.fi}, W.F.
Chen$^2$\renewcommand{\thefootnote}{\ddagger}\footnote{E-mail:
wchen@msnet.mathstat.uoguelph.ca} and R. Gonz\'{a}lez
Felipe$^3$\renewcommand{\thefootnote}{\ast}\footnote{E-mail:
gonzalez@gtae3.ist.utl.pt}}

\address{$^1$ High Energy Physics Division, Department of Physics,
University of Helsinki\\
and Helsinki Institute of Physics, FIN-00014, Helsinki, Finland\\
$^2$ Department of Mathematics and Statistics, University of
Guelph\\ Guelph, Ontario, Canada N1G 2W1\\
$^3$ Centro de F\'{\i}sica das Interac\c{c}\~{o}es Fundamentais, Departamento de
F\'{\i}sica\\ Instituto Superior T\'{e}cnico, 1049-001 Lisboa, Portugal}

\maketitle

\begin{abstract}
The Schwinger proper-time method is an effective calculation method,
explicitly gauge invariant and nonperturbative. We make use of this
method to investigate the radiatively induced Lorentz and
$CPT$-violating effects in quantum electrodynamics when an axial
vector interaction term is introduced in the fermionic sector. The
induced Lorentz and $CPT$-violating Chern-Simons term coincides with
the one obtained using a covariant derivative expansion but differs
from the result usually obtained in other regularization schemes. A
possible ambiguity in the approach is also discussed.\\

\vspace{5mm} \noindent \emph{PACS}: 11.30.Cp, 11.30.Er, 12.20.-m,
11.10.Gh


\end{abstract}

\vspace{4ex}

The question of inducing a Lorentz- and $CPT$-violating Chern-Simons
(CS) term through radiative corrections by adding a term of the form
$\overline{\psi}\gamma_5\bslash\psi$ ($b_\mu$ is a prescribed
constant 4-vector) to the conventional QED Lagrangian, has been
recently addressed by several authors
\cite{carroll,colladay,coleman,jackiw,chung,jackiw2,chen,victoria,chan}.
A theoretical investigation on the physical effects of Lorentz and
$CPT$ violation in electrodynamics has been made by the inclusion of
a CS term $\frac{1}{2} \epsilon^{\mu\nu\lambda\rho} k_\mu
F_{\nu\lambda} A_\rho$ to the Lagrangian density
\cite{carroll,colladay}. The constant 4-vector $k_\mu$ selects a
preferred direction in space-time, thus violating both Lorentz
invariance and the discrete $CPT$ symmetry. Since up to now the
observation of distant galaxies does not give any experimental
evidence for the occurrence of such an effect, the vector $k_\mu$ is
severely constrained and should effectively vanish
\cite{goldhaber,jackiw3}. On the other hand, one may ask the question
whether such a term can be induced through radiative corrections when
Lorentz and $CPT$ symmetries are violated in other sectors of the
theory, for instance in the fermionic one. In this case the stringent
bounds on $k_\mu$ would translate into constraints on these sectors.

The quantum electrodynamics with a Lorentz- and $CPT$-violating axial
vector interaction term in the fermionic sector is described by the
Lagrangian density
\begin{equation}
{\cal L}=\overline{\psi} \left(i\partialslash-e\Aslash-m -
\gamma_5\bslash\right)\psi \ . \label{eq1}
\end{equation}
The full quantum effective action is obtained by integrating out the
fermionic fields,
\begin{eqnarray}
Z[A]&=& e^{i\Gamma_{\rm eff}}=\int {\cal D}\overline{\psi} {\cal D}
\psi e^{i\int d^4x {\cal L}}= \det \left(i\partialslash-e\Aslash-m
-\gamma_5\bslash\right),\nonumber\\
\Gamma_{\rm eff}&=&-i\Tr\ln \left(i\partialslash
-e\Aslash-m-\gamma_5\bslash\right)\ , \label{eq2}
\end{eqnarray}
where the trace is taken over both space-time coordinates and spinor
indices.

Several calculations have been carried out to determine the induced
CS term, which is proportional to $b_\mu$. While nonperturbative
treatments in $b_\mu$ in the vacuum polarization tensor
\cite{jackiw,chung,victoria} and the covariant derivative expansion
\cite{chan} yield definite and nonzero values for the CS coefficient,
the perturbative treatment gives an ambiguous result, which depends
on the choice of the regularization scheme \cite{colladay,coleman}.
Indeed, if one chooses Pauli-Villars regularization to enforce gauge
invariance for all axial momenta, then the induced mass-independent
CS coefficient will vanish \cite{colladay}. If instead dimensional
regularization is used, then different answers can be obtained
depending on how the matrix $\gamma_5$ is generalized to arbitrary
dimensions. A universal ambiguous result has also been given in terms
of differential regularization \cite{chen}. Even in terms of the
nonperturbative treatment in $b_\mu$, the calculations based on the
Feynman diagram expansion \cite{jackiw,chung,victoria} and the
non-Feynman diagram method (e.g. the covariant derivative expansion)
\cite{chan} yield different results.

The purpose of this work is to calculate the Lorentz and
$CPT$-violating Chern-Simons term radiatively induced by the axial
coupling of the vector $b_\mu$ in the effective action (\ref{eq2}).
We shall employ the well-known Schwinger proper-time method
\cite{schwinger}, which is not only a gauge invariant procedure, but
also allows us to obtain an explicit solution of the equations of
motion for the special case of gauge fields which produce a constant
field strength tensor $F_{\mu\nu}$. Recently, this approach was
employed in combination with the analytic regularization method to
investigate the radiatively induced Lorentz and CPT breaking in a
constant external electromagnetic field \cite{tomazelli}. It was
shown that the induced CS coefficient exhibits a logarithmic
contribution, but this result explicitly contradicts the exact
calculation performed in Ref. \cite{victoria}. As we shall see below,
if one follows the original route proposed by Schwinger
\cite{schwinger}, such a nonperturbative and gauge-invariant
formulation leads to an induced CS current with a finite and definite
nonzero value, which is identical to the result obtained in the
covariant derivative expansion \cite{chan} but differs from the
result usually obtained in the calculation based on the Feynman
diagram technique \cite{jackiw,chung,victoria}.

Although the evaluation of the effective action (\ref{eq2}) is not
unique and crucially depends on the regularization scheme one adopts,
it seems to us that the discrepancies in the results obtained for the
CS term are not fully due to the chosen regularization. To make our
point clear, let us first analyze the properties of a regularization
dependent ambiguity. As is well known, a regularization is a
temporary modification of the original theory and is a necessary
procedure to extract out the divergences so that renormalization can
be implemented. Different regularization schemes actually provide
different methods to evaluate a quantum correction. For the finite
terms in the quantum corrections, the role of a gauge invariant
regularization is no longer regulating the divergent quantities but
rather imposing gauge invariance \cite{chan}. One typical example is
provided by the chiral anomaly, which arises in any vector gauge
invariant regularization scheme such as Pauli-Villars regularization
and dimensional regularization, but vanishes in any non-invariant
regularization method. The reason is that these regularization
methods preserve the vector gauge symmetry but, simultaneously,
violate the axial vector gauge symmetry of the classical theory. On
the other hand, differential regularization clearly shows that the
anomaly manifests itself in both the vector and axial vector gauge
symmetries. This regularization method does not impose any symmetry
\emph{a priori} and to preserve any desired symmetry in the theory
one then chooses the (undetermined) renormalization scale at the
final stage of the calculation \cite{johnson,chen}. However, all the
regularization schemes which preserve the same symmetry should give
the same finite result. If different results are produced by such
regularization methods, there must be certain inconsistency in the
calculation procedure.

Moreover, it has recently been claimed \cite{chan} that if one uses a
covariant derivative expansion to calculate the coefficient of the
radiatively induced CS term, then there is a discrepancy with the
results previously obtained \cite{jackiw,chung,victoria} by using the
widely accepted approach of the chiral anomaly \cite{jackiw}. It was
explicitly shown in Ref. \cite{chan} that when compared with the
standard perturbative expansion in powers of the gauge field $A_\mu$,
the one-loop vacuum polarization tensor in the covariant derivative
expansion has an additional term, whose surface term gives a
non-vanishing contribution to the usual result and which is precisely
the source of the discrepancy. Since the essence of the covariant
derivative expansion is to develop the local effective Lagrangian in
a series of powers of $\Pi_\mu=i\partial_\mu - e A_\mu$, rather than
$i\partial_\mu $ and $A_\mu$ separately \cite{chan1,gail}, this
formalism is explicitly gauge covariant. In this sense the covariant
derivative expansion and the Schwinger proper-time method share the
same features, and this is another motivation for us to employ the
latter method in our calculation.

Let us first write the trace in Eq.\,(\ref{eq2}) in the following
equivalent form,
\begin{equation}
\Tr\ln \left(i\partialslash-e\Aslash-m -\gamma_5 \bslash\right)
= \Tr\ln\left(i\partialslash-e\Aslash-m\right) -\int_0^1dz
\Tr\left(\frac{1}{i\partialslash-e\Aslash-m -z\gamma_5 \bslash}
\gamma_5\bslash\right). \label{eq3}
\end{equation}
Then the quantum effective action can be rewritten as
\begin{equation}
\Gamma_{\rm eff}=\Gamma_{\rm eff}^{(0)}+\Gamma_{\rm eff}^{(1)}\ ,
\label{eq4}
\end{equation}
where
\begin{eqnarray}
\Gamma_{\rm eff}^{(0)}&=&-i\Tr \ln \left(i\partialslash-e\Aslash
-m\right),\nonumber\\
\Gamma_{\rm eff}^{(1)}&=&i\int_0^1dz
\Tr\left(\frac{1}{i\partialslash-e\Aslash-m -z\gamma_5 \bslash}
\gamma_5\bslash\right). \label{eq5}
\end{eqnarray}
The radiatively induced CS term comes only from the $\Gamma_{\rm
eff}^{(1)}$ part. Moreover, since $\Gamma_{\rm CS} \propto b_\mu$ we
can neglect the dependence of $b$ in the denominator of the trace in
$\Gamma_{\rm eff}^{(1)}$. Thus we have,
\begin{equation}
\Gamma_{\rm CS}=i\Tr \left(\frac{1}{i\partialslash-e\Aslash-m} \gamma_5
\bslash\right) =i\Tr\left(\frac{1}{\Pislash-m} \gamma_5 \bslash\right)\ ,
\label{eq6}
\end{equation}
with the commutation relations $[x_\mu,\Pi_\nu]=-ig_{\mu\nu}\ ,\
[\Pi_\mu,\Pi_\nu]=-ieF_{\mu\nu}$.

To calculate the radiatively induced CS term we make use of the
Schwinger proper-time method \cite{schwinger}. We begin by writing
Eq.\,(\ref{eq6}) as
\begin{equation}
\Gamma_{\rm CS} = i\int d^4 x\Tr\left(\langle x|
\frac{1}{\Pislash-m}| x\rangle\gamma_5 \bslash \right)
= \left. i\int d^4x \Tr\left[G(x,x^{\prime})\gamma_5
\bslash \right]\right|_{x^{\prime}\rightarrow x}\ , \label{eq7}
\end{equation}
where the function $G(x,x^{\prime})$ is defined by
\begin{equation}
\left(\Pislash-m\right)G(x,x^{\prime})=\delta^{(4)}(x-x^{\prime})\ ,
\label{eq8}
\end{equation}
and the limit $x^{\prime} \rightarrow x$ is performed by taking the
average of the terms obtained by letting $x^{\prime} \rightarrow x$
from the future and from the past \cite{schwinger}. In operator form,
\begin{equation}
G(x,x^{\prime})=\langle x|G|x^{\prime}\rangle,
~~G=\frac{1}{\Pislash-m} =\left({\Pislash
+m}\right)\frac{1}{\Pislash^2 -m^2}= \frac{1}{\Pislash^2 -m^2}
\left({\Pislash +m}\right). \label{eq9}
\end{equation}
Therefore we have two equivalent integral representations for the
operator $G$,
\begin{equation}
G=-i\left(\Pislash+m\right)\int_{-\infty}^0 ds\, e^{im^2s}\, U(s)
=-i\int_{-\infty}^0 ds\, e^{im^2s}\, U(s)\, \left(\Pislash+m\right),
 \label{eq10}
\end{equation}
where $U(s) \equiv e^{-i{\cal H}s}$ and
\begin{equation}
{\cal H}=\Pislash^2=\Pi^2-\frac{e}{2}\sigma\cdot F, ~~ \sigma\cdot
F=\sigma_{\mu\nu}F^{\mu\nu},~~\sigma_{\mu\nu}=
\frac{i}{2}\left[\gamma_\mu,\gamma_\nu\right]\, . \label{eq11}
\end{equation}
Eqs.\,(\ref{eq9}) and (\ref{eq10}) show that the calculation of
$G(x,x^{\prime})$ lies in the evaluation of
\begin{equation}
\langle x^{\prime} |U(s)|x^{\prime\prime}\rangle = \langle
x^{\prime}(s)|x^{\prime\prime}(0)\rangle\ . \label{eq12}
\end{equation}
The operator $U(s)$ can be regarded as an evolution operator of a
system governed by the ``Hamiltonian" ${\cal H}$ in the ``time" $s$.
The Heisenberg equations of motion for the operators $x_\mu (s)$ and
$\Pi_\mu (s)$ read
\begin{eqnarray}
i\frac{dx_\mu}{ds}&=& [x_\mu,{\cal H}]=-2i\Pi_\mu,\nonumber\\
i\frac{d\Pi_\mu}{ds}&=& [\Pi_\mu,{\cal H}]= -ie
\left(F_{\mu\nu}\Pi^{\nu}+\Pi^{\nu}F_{\mu\nu}\right)
-\frac{1}{2}ie\sigma^{\nu\rho}\partial_{\mu}F_{\nu\rho} \nonumber\\
&=&-2ie F_{\mu\nu}\Pi^{\nu}+e\partial^\nu F_{\mu\nu}
-\frac{1}{2}ie\sigma^{\nu\rho}\partial_\mu F_{\nu\rho}. \label{eq13}
\end{eqnarray}
Correspondingly, the transformation function satisfies the
differential equations
\begin{eqnarray}
&&i\frac{d}{ds}\langle x^{\prime}(s)|x^{\prime\prime}(0)\rangle
=\langle x^{\prime}(s)|{\cal H}|x^{\prime\prime}(0)\rangle, \nonumber\\
&&\left[i\frac{\partial}{\partial x^{\prime\mu}} -eA_\mu
(x^{\prime})\right]\langle x^{\prime}(s)|x^{\prime\prime}(0)\rangle
=\langle x^{\prime}(s)|\Pi_\mu (s)|x^{\prime\prime}(0)\rangle, \nonumber\\
&&\left[-i\frac{\partial}{\partial x^{\prime\prime\mu}} -eA_\mu
(x^{\prime\prime})\right]\langle x^{\prime}(s)
|x^{\prime\prime}(0)\rangle =\langle x^{\prime}(s)|\Pi_\mu
(0)|x^{\prime\prime}(0)\rangle, \label{eq14}
\end{eqnarray}
with the initial condition
\begin{equation}
\left. \langle x^{\prime}(s)|x^{\prime\prime}(0)\rangle \right|_{s
\rightarrow 0}= \delta^{(4)}(x^{\prime}-x^{\prime\prime}). \label{eq15}
\end{equation}

We shall now solve the Heisenberg equations (\ref{eq13}) for the
special case of a constant field strength tensor, i.e. we choose
$F_{\mu\nu}$ to be constant, $A_{\mu}=-F_{\mu\nu}x^{\nu}/2$. This
will allow us to calculate the transformation function $\langle
x^{\prime}(s)|x^{\prime\prime}(0)\rangle$ in an exact analytical
form. In this case the proper-time dynamical equations (\ref{eq13})
are simplified to
\begin{equation}
\frac{dx_\mu}{ds}=-2\Pi_\mu\ , ~~\frac{d\Pi_\mu}{ds}
=-2eF_{\mu\nu}\Pi^\nu\ ,
\label{eq16}
\end{equation}
or in matrix notation,
\begin{equation}
\frac{dx}{ds}=-2\Pi\ , ~~~\frac{d\Pi}{ds}=-2eF\Pi\ . \label{eq17}
\end{equation}
The formal solutions can be easily written out,
\begin{equation}
\Pi(s)=e^{-2eFs}\Pi(0)\ , ~~~
x(s)-x(0)=\frac{1}{eF}\left[e^{-2eFs}-1\right]\Pi(0)\ . \label{eq18}
\end{equation}

To get the transformation function, we express $\Pi(s)$ in terms
of $x(s)$,
\begin{eqnarray}
\Pi(0)&=&-\frac{1}{2}eF e^{eFs} \sinh^{-1}(eFs)
\left[x(s)-x(0)\right], \nonumber\\
\Pi(s)&=&-\frac{1}{2}eF e^{-eFs} \sinh^{-1}(eFs)
\left[x(s)-x(0)\right] \nonumber\\
&=&-\left[x(s)-x(0)\right]\frac{1}{2}eF e^{eFs} \sinh^{-1}(eFs)\ .
\label{eq19}
\end{eqnarray}
Therefore,
\begin{eqnarray}
\Pi^2(s)&=& \left[x(s)-x(0)\right]K\left[x(s)-x(0)\right], \nonumber\\
K&=&\frac{1}{4}e^2F^2\sinh^{-2}(eFs). \label{eq20}
\end{eqnarray}
Using the commutation relation
\begin{eqnarray}
\left[x(s),x(0)\right]&=&\left[x(0)+(eF)^{-1}\left(e^{-2eFs}-1
\right)\Pi(0), x(0)\right]\nonumber\\
&=&i(eF)^{-1}\left(e^{-2eFs}-1\right), \label{eq21}
\end{eqnarray}
we can write
\begin{equation}
\Pi^2(s) = x(s)K x(s)-2 x(s)K x(0)+x(0) K x(0)
-\frac{i}{2}\Tr\left[eF\coth\left(eFs\right)\right]\ , \label{eq22}
\end{equation}
where the trace extends over space-time indices.

Taking into account Eq.\,(\ref{eq22}), the first differential
equation in Eq.\,(\ref{eq14}) becomes
\begin{eqnarray}
i\frac{d}{ds}\langle x^{\prime}(s)|x^{\prime\prime}(0)\rangle
&=&\left\{-\frac{e}{2}\sigma\cdot F+(x^{\prime}-x^{\prime\prime})
K(x^{\prime}-x^{\prime\prime})\right.\nonumber\\
&&\left.-\frac{i}{2}\Tr\left[eF\coth\left(eFs\right)\right] \right\}
\langle x^{\prime}(s)|x^{\prime\prime}(0)\rangle, \label{eq23}
\end{eqnarray}
which has the solution
\begin{equation}
\langle x^{\prime}(s)|x^{\prime\prime}(0)\rangle = C(x^{\prime},
x^{\prime\prime}) e^{-L(s)}s^{-2}e^{\frac{i}{4}
(x^{\prime}-x^{\prime\prime})eF\coth(eFs)
(x^{\prime}-x^{\prime\prime})}e^{\frac{i}{2}e\sigma\cdot Fs}\ ,
\label{eq24}
\end{equation}
where
\begin{equation}
 L(s)=\frac{1}{2}\Tr\ln\left[(eFs)^{-1}\sinh(eFs) \right].
\label{eq24a}
\end{equation}
The coefficient function $C(x^{\prime}, x^{\prime\prime})$ can be
determined from the second and third differential equations in
(\ref{eq14}). Since from Eq.\,(\ref{eq19}) we have,
\begin{eqnarray}
\langle x^{\prime}(s)|\Pi(0)|x^{\prime\prime}(0)\rangle
&=&-\frac{1}{2}\left[eF\coth \left(eFs\right)+eF\right]
(x^{\prime}-x^{\prime\prime}) \langle
x^{\prime}(s)|x^{\prime\prime}(0)\rangle\ ,\nonumber\\
\langle x^{\prime}(s)|\Pi(s)|x^{\prime\prime}(0)\rangle
&=&-\frac{1}{2} \left[eF\coth\left(eFs\right)-eF\right]
(x^{\prime}-x^{\prime\prime}) \langle
x^{\prime}(s)|x^{\prime\prime}(0)\rangle\ , \label{eq25}
\end{eqnarray}
then Eqs.\,(\ref{eq14}), (\ref{eq24}) and (\ref{eq25}) lead to the
following differential equations for $C(x^{\prime},
x^{\prime\prime})$:
\begin{eqnarray}
&&\left[i\frac{\partial}{\partial x^{\prime \mu}} -eA_\mu
(x^{\prime})-\frac{e}{2}
F_{\mu\nu}(x^{\prime}-x^{\prime\prime})^{\nu}\right] C(x^{\prime},
x^{\prime\prime})=0\ ,\nonumber\\
&&\left[-i\frac{\partial}{\partial x^{\prime\prime\mu}} -eA_\mu
(x^{\prime\prime})+\frac{e}{2}
F_{\mu\nu}(x^{\prime}-x^{\prime\prime})^{\nu}\right] C(x^{\prime},
x^{\prime\prime})=0\ . \label{eq26}
\end{eqnarray}
The solution of these equations is given by
\begin{equation}
C(x^{\prime}, x^{\prime\prime}) = C \Phi(x^{\prime},
x^{\prime\prime})\ ,\quad \Phi(x^{\prime}, x^{\prime\prime})=\exp
\left[-ie\int_{x^{\prime\prime}}^{x^{\prime}} dx^{\mu}
A_{\mu}(x)\right]\ , \label{eq27}
\end{equation}
where we have chosen the integration path to be a straight line
connecting $x^{\prime}$ and $x^{\prime\prime}$, since the integral in
the exponent of Eq.\,(\ref{eq27}) is independent of the path. The
overall constant $C$ can be fixed by the initial condition
(\ref{eq15}) and we find
\begin{equation}
C=-\frac{i}{(4\pi)^2}\ . \label{eq29}
\end{equation}
Thus we finally get the transformation function
\begin{eqnarray}
\langle x^{\prime}(s)|x^{\prime\prime}(0)\rangle
&=&-\frac{i}{(4\pi)^2}\Phi(x^{\prime}, x^{\prime\prime})
e^{-L(s)}s^{-2}e^{\frac{i}{4} (x^{\prime}-x^{\prime\prime})
eF\coth(eFs)(x^{\prime}-x^{\prime\prime})}
e^{\frac{i}{2}e\sigma\cdot Fs}\ , \label{eq30}
\end{eqnarray}
and the corresponding Green function
\begin{eqnarray}
G(x^{\prime},x^{\prime\prime}) &=& -i\int_{-\infty}^0 ds\, e^{im^2s}
\left[\gamma^{\mu}\langle x^{\prime}(s)|\Pi_\mu
(s)|x^{\prime\prime}(0)\rangle +m \langle
x^{\prime}(s)|x^{\prime\prime}(0)\rangle\right]\nonumber\\
&=&-i\int_{-\infty}^0 ds\, e^{im^2s} \left[\langle
x^{\prime}(s)|\Pi_\mu(0)|x^{\prime\prime}(0)\rangle \gamma^{\mu} +m
\langle x^{\prime}(s)|x^{\prime\prime}(0)\rangle\right] \
.\label{eq31}
\end{eqnarray}

We are now in position to obtain the radiatively induced CS term
(\ref{eq7}). Using the trace relation $\Tr
\left[\gamma_5\gamma_{\mu_1}\cdots\gamma_{\mu_{2n+1}}\right] = 0$, we
have
\begin{eqnarray}
\Tr\left[G(x,x^{\prime})\gamma_5\bslash \right]
&=&-i\Tr\left[\gamma_5\bslash\gamma^{\mu}
 \int_{-\infty}^0 ds\, e^{im^2s}
\langle x(s)|\Pi_\mu(s)|x^{\prime}(0)\rangle \right]\nonumber\\
&=&\Tr\left[\left(-i\gamma_5
b^\mu+\gamma_5\sigma^{\mu\nu}b_\nu\right) \int_{-\infty}^0 ds\,
e^{im^2s} \langle x(s)|\Pi_\mu(s)|x^{\prime}(0) \rangle
\right]\nonumber\\
&=&-i\Tr\left[\gamma^{\mu} \gamma_5\bslash \int_{-\infty}^0 ds\,
e^{im^2s}\langle x(s)| \Pi_\mu(0)|x^{\prime}(0)
\rangle \right]\nonumber\\
&=&\Tr\left[\left(i\gamma_5b^\mu+\gamma_5\sigma^{\mu\nu}b_\nu \right)
\int_{-\infty}^0 ds\, e^{im^2s}\langle x(s)| \Pi_\mu(0)|x^{\prime}(0)
\rangle\right]. \label{eq33}
\end{eqnarray}
On averaging these two equivalent expressions we find that
\begin{eqnarray}
\Tr\left[G(x,x^{\prime})\gamma_5\bslash\right]
&=&-ib^{\mu}\Tr\left\{\gamma_5 \int_{-\infty}^0 ds\, e^{im^2s}\langle
x(s)|\frac{1}{2}\left[\Pi_\mu(s)- \Pi_\mu(0)\right]|x^{\prime}(0)
\rangle\right\}\nonumber\\
&+&b_\nu \Tr\left\{\gamma_5\sigma^{\mu\nu} \int_{-\infty}^0 ds\,
e^{im^2s}\langle x(s)|\frac{1}{2}\left[\Pi_\mu(s)+
\Pi_\mu(0)\right]|x^{\prime}(0)\rangle \right\}\ . \label{eq34}
\end{eqnarray}
According to Eqs.\,(\ref{eq25}),
\begin{eqnarray}
\langle x(s)|\frac{1}{2}\left[\Pi_\mu(s)-
\Pi_\mu(0)\right]|x^{\prime}(0)\rangle &=&
\frac{e}{2}F_{\mu\nu}(x-x^{\prime})^{\nu} \langle
x(s)|x^{\prime}(0)\rangle\ , \nonumber\\
\langle x(s)|\frac{1}{2}
\left[\Pi_\mu(s)+\Pi_\mu(0)\right]|x^{\prime}(0)\rangle &=&
-\frac{e}{2}F_{\mu\alpha}\left[\coth(eFs) \right]^{\alpha\beta}
(x-x^{\prime})_{\beta}\langle x(s)|x^{\prime}(0)\rangle\ .
\label{eq35}
\end{eqnarray}
Thus we obtain
\begin{eqnarray}
\Tr\left[G(x,x^{\prime})\gamma_5\bslash\right] &=&-\frac{e}{2}b_\nu
F_{\mu\alpha}\Tr \left[\gamma_5\sigma^{\mu\nu}\int_{-\infty}^0 ds\,
e^{im^2s} \left[\coth(eFs)\right]^{\alpha\beta}(x-x^{\prime})_{\beta}
\langle x(s)|x^{\prime}(0)\rangle\right] \nonumber\\
&-&\frac{ie}{2}b_{\mu}F^{\mu\nu}(x-x^{\prime})_{\nu}
\Tr\left[\gamma_5\int_{-\infty}^0 ds\, e^{im^2s} \langle
x(s)|x^{\prime}(0)\rangle\right]\ . \label{eq36}
\end{eqnarray}
Using Eq.\,(\ref{eq30}) we can write then
\begin{eqnarray}
\Tr\left[G(x,x^{\prime})\gamma_5\bslash\right] &=& \frac{ie}{32\pi^2}
\Phi(x,x^{\prime}) b_\nu \int_{-\infty}^0 \frac{ds}{s^2} \ e^{im^2s}
e^{-L(s)}e^{\frac{i}{4} (x-x^{\prime})eF\coth(eFs)
(x-x^{\prime})} \nonumber\\
&\times& \left\{ F_{\mu\alpha} \Tr\left[\gamma_5 \sigma^{\mu\nu}
e^{\frac{i}{2}e\sigma\cdot Fs}\right] [\coth(eFs)]^{\alpha\beta}
(x-x^\prime)_\beta \right. \nonumber\\
&-& \left. iF^{\mu\nu}(x-x^\prime)_\mu \Tr\left[\gamma_5
e^{\frac{i}{2}e\sigma\cdot Fs}\right] \right\}\ . \label{eq36a}
\end{eqnarray}

It turns out to be more convenient to calculate the ground-state
current in the presence of the background field $A_\mu(x)$,
\begin{equation}
\langle J_{\mu}(x) \rangle =\frac{\delta \Gamma_{\rm CS}}{\delta
A_{\mu}(x)}\ . \label{eq37}
\end{equation}
From Eqs.\,(\ref{eq7}) and (\ref{eq36a}), and after deforming the
integration path by the substitution $s \rightarrow -is$ we get the
nonperturbative expression for the CS current:
\begin{eqnarray}
\langle J_{\mu}(x) \rangle &=& \frac{ie^2}{32\pi^2}
(x-x^{\prime})_\mu \Phi(x,x^{\prime})
b_\nu\int_{-\infty}^0\frac{ds}{s^2}\ e^{m^2s} e^{-L(s)}
e^{-\frac{1}{4}(x-x^{\prime})eF\cot(eFs)(x-x^{\prime})} \nonumber\\
&\times&\left\{-F_{\rho\alpha} \Tr\left[\gamma_5\sigma^{\rho\nu}
e^{\frac{e}{2}\sigma\cdot Fs}\right]
\left[\cot(eFs)\right]^{\alpha\beta}(x-x^{\prime})_{\beta}\right. \nonumber\\
&+& \left. \left. F^{\rho\nu}
(x-x^{\prime})_\rho \Tr\left[\gamma_5 e^{\frac{e}{2}\sigma\cdot Fs}
\right]
\right\}\right|_{x^{\prime}{\rightarrow}x}\ , \label{eq38}
\end{eqnarray}
where the operator $L(s)$ is given by Eq.\,(\ref{eq24a}) after the
corresponding substitution $s \rightarrow -is$.

To evaluate the CS current, the following operations should be
performed. First we notice that since the gauge-invariant quantity
$e^{-L(s)}$ is a determinant, it can be evaluated from the
eigenvalues of the matrix $F$, which can be easily found with the
assistance of the relations:
\begin{eqnarray}
&&F^{\mu\rho}\widetilde{F}_{\rho\nu}=-\delta^{\mu}_{\nu}
{\cal G},\label{eq39}\\
&&\widetilde{F}_{\mu\rho}\widetilde{F}^{\rho\nu}
-F_{\mu\rho}F^{\rho\nu} =2\delta_{\mu}^{~\nu}{\cal F}, \label{eq40}
\end{eqnarray}
where $\widetilde{F}^{\mu\nu}
=\frac{1}{2}\epsilon^{\mu\nu\lambda\rho} F_{\lambda\rho}$ is the dual
field strength tensor, ${\cal F}=\frac{1}{4}F_{\mu\nu}F^{\mu\nu}
=-\frac{1}{4}\widetilde{F}_{\mu\nu}\widetilde{F}^{\mu\nu} =
-\frac{1}{2}({\bf E}^2-{\bf B}^2)$ and ${\cal G}
=\frac{1}{4}F_{\mu\nu}\widetilde{F}^{\mu\nu}=-{\bf E}\cdot{\bf B}$.
We iterate the eigenvalue equations:
\begin{equation}
F_{\mu\nu}\psi^{\nu}=\lambda\psi_\mu\ , \quad
\widetilde{F}_{\mu\nu}\psi^{\nu} = -\frac{1}{\lambda}{\cal
G}\psi_\mu\ , \label{eq42}
\end{equation}
to obtain
\begin{equation}
F_{\mu\rho}F^{\rho\nu}\psi_\nu=\lambda^2\psi_\mu, \quad
\widetilde{F}_{\mu\rho}\widetilde{F}^{\rho\nu}\psi_\nu
=\frac{1}{\lambda^2}{\cal G}^2\psi_\mu\ . \label{eq43}
\end{equation}
The identity (\ref{eq40}) yields then the equation
\begin{equation}
\lambda^4+2{\cal F}\lambda^2-{\cal G}^2=0\ , \label{eq44}
\end{equation}
which has four solutions, $\pm \lambda_1$ and  $\pm \lambda_2$,
\begin{equation}
\lambda_{1,2}=\frac{i}{\sqrt{2}}\left[\sqrt{{\cal F}+i{\cal G}} \pm
\sqrt{{\cal F}-i{\cal G}}\right]\ . \label{eq45}
\end{equation}
The quantity $e^{-L(s)}$ can now be expressed in terms of these
eigenvalues,
\begin{equation}
e^{-L(s)}=(es)^2\frac{\lambda_1}{\sin(e\lambda_1s)}
\frac{\lambda_2}{\sin(e\lambda_2s)} =\frac{(es)^2 {\cal G}}{\im[\cosh
esX]}\ , \label{eq46}
\end{equation}
where $X=\sqrt{2({\cal F}+i{\cal G})}$.

Next we need to evaluate the Dirac traces. It is easy to show that
the following matrix decomposition holds:
\begin{equation}
e^{\frac{e}{2}\sigma\cdot Fs}=c_1 \openone+c_2\sigma\cdot
F+ic_3\gamma_5+ic_4\sigma\cdot F\gamma_5\ , \label{eq48}
\end{equation}
where the coefficients $c_i$ are given by
\begin{eqnarray}
c_1 &=& \re \cosh(esX)\ ,\quad \quad \quad \ c_3 = \im \cosh(esX)\
,\nonumber\\
c_2 &=&\re\left[\sinh(esX)/2X\right]\ ,\quad c_4 =
\im\left[\sinh(esX)/2X\right]\ .\nonumber\\
\end{eqnarray}
In the expression (\ref{eq38}) for the CS current the following Dirac
traces appear:
\begin{eqnarray}
\Tr\left\{\gamma_5 e^{\frac{e}{2}\sigma\cdot Fs}\right\}
&=& 4ic_3\ , \label{eq50} \\
 \Tr\left\{\gamma_5 \sigma^{\rho\nu} e^{\frac{e}{2}\sigma\cdot
Fs}\right\} &=& 8i\left(c_2 \widetilde{F}^{\rho\nu} +c_4
F^{\rho\nu}\right) \, . \label{eq51}
\end{eqnarray}
In deriving the above formulas we have used the relations:
\begin{eqnarray}
& &\Tr\ \gamma_5 = \Tr\ \sigma_{\mu\nu} = \Tr\{\gamma_5
\sigma_{\mu\nu}\} = 0\ , \nonumber\\
& &\Tr\{\sigma_{\rho\nu}\sigma_{\alpha\beta}\} =
4(g_{\rho\alpha}g_{\nu\beta}-g_{\rho\beta}g_{\nu\alpha})\ ,
\nonumber\\
& & \Tr\{\gamma_5\sigma^{\rho\nu}\sigma^{\alpha\beta}\} =
4i\epsilon^{\rho\nu\alpha\beta}\ .\label{eq52}
\end{eqnarray}
Substituting Eqs.\,(\ref{eq46}), (\ref{eq50}) and (\ref{eq51}) into
Eq.\,(\ref{eq38}) we obtain
\begin{eqnarray}
\langle J_{\mu}(x) \rangle &=& \frac{e^4}{4\pi^2}\, (x-x^\prime)_\mu
\Phi(x,x^\prime) b_\nu {\cal G} \int_0^{\infty}ds \,
\frac{e^{-m^2s}}{\im[\cosh(esX)]}\,
e^{\frac{1}{4}(x-x^{\prime})eF\cot(eFs)(x-x^{\prime})}\nonumber\\
&\times& \left\{F_{\rho\alpha} \left(c_2
\widetilde{F}^{\rho\nu}+c_4 F^{\rho\nu}\right)
\left[\cot(eFs)\right]^{\alpha\beta} (x-x^{\prime})_{\beta} -
\left.\frac{c_3}{2}F^{\rho\nu} (x-x^{\prime})_\rho
\right\}\right|_{x^{\prime}{\rightarrow}x}\, . \label{eq53}
\end{eqnarray}
For our purposes it is sufficient to evaluate Eq.\,(\ref{eq53}) in
the weak field approximation. In this case
\begin{eqnarray}
&&\im[\cosh(esX)] \simeq (es)^2 {\cal G}\ ,\quad eF\cot(eFs) \simeq
\frac{1}{s}\ , \nonumber\\
&&c_2 \simeq \frac{es}{2}\ ,\quad c_3 \simeq 0\ ,\quad c_4\simeq 0\ ,
\label{eq54}
\end{eqnarray}
and therefore the leading term in the CS current (\ref{eq53}) reads
as
\begin{equation}
\langle J_{\mu}(x) \rangle = \left. \frac{e^2}{2\pi^2}
\Phi(x,x^\prime) \frac{m^2 K_1\left(\left[-m^2(
x-x^{\prime})^2\right]^{1/2}\right)}{\left[-m^2(
x-x^{\prime})^2\right]^{1/2}}(x-x^{\prime})_\mu (x-x^{\prime})_\rho
b_{\nu} \widetilde{F}^{\rho\nu}\right|_{x^\prime\rightarrow x}\ ,
\label{eq56}
\end{equation}
where $K_1(z)$ is the first-order modified Bessel function defined
through its integral representation
$2\sqrt{z}\,K_1(2\sqrt{z})=\int_0^{\infty} ds\ s^{-2}\ e^{-sz-1/s}$.
Finally, making use of the following formulae
\begin{eqnarray}
&&\Phi (x,x^\prime)|_{x^\prime\rightarrow x}=1\ , \quad
\left.\frac{K_1\left(\left[-m^2(
x-x^{\prime})^2\right]^{1/2}\right)}{\left[-m^2(
x-x^{\prime})^2\right]^{1/2}}\right|_{x^\prime\rightarrow x} = -
\frac{1}{m^2 (x-x^\prime)^2}\ ,\nonumber\\
&&\nonumber\\
&&\left.\frac{(x-x^\prime)_{\mu}(x-x^\prime)_{\nu}}{(x-x^\prime)^2}
\right|_{x^\prime\rightarrow x} \equiv \left. \frac{1}{\int
d^4x^\prime}\int d^4x^\prime
\frac{(x-x^\prime)_{\mu}(x-x^\prime)_{\nu}}{(x-x^\prime)^2}
\right|_{x^\prime\rightarrow x} = \frac{1}{4}\,g_{\mu\nu}\ ,
\label{eq58}
\end{eqnarray}
we obtain
\begin{equation}
\langle J_{\mu}(x) \rangle
=-\frac{e^2}{8\pi^2}\widetilde{F}^{\mu\nu} b_\nu =-
\frac{e^2}{16\pi^2}\epsilon^{\mu\nu\lambda\rho}b_\nu
F_{\lambda\rho}\ . \label{eq59}
\end{equation}
By functionally integrating over the gauge field $A_\mu$ (cf.
Eq.\,(\ref{eq37})) we get the radiatively induced Lorentz and $CPT$
violating CS action
\begin{equation}
\Gamma_{\rm CS}=\frac{e^2}{16\pi^2}\int d^4 x
\epsilon^{\mu\nu\lambda\rho} b_{\mu}A_{\nu}F_{\lambda\rho}\ .
\label{eq60}
\end{equation}

Eq.\,(\ref{eq60}) shows that the Schwinger proper-time formulation
yields a finite and nonzero value for the radiatively induced CS
term. We note that this result coincides with the one obtained in
Ref. \cite{chan} using a covariant derivative expansion, but it
differs from the result of Refs. \cite{jackiw,chung,victoria}, where
a coefficient $3/(32\pi^2)$ is obtained instead of $1/(16\pi^2)$. As
discussed before, this discrepancy is not due to the ambiguity in the
choice of a regularization scheme. The covariant derivative expansion
shows clearly that the vacuum polarization tensor receives a
non-Feynman diagram contribution \cite{chan}, which is the origin of
the discrepancy in the results for the Chern-Simons coefficient. The
Schwinger constant field approximation method has a common feature
with the covariant derivative expansion, namely, the (axial vector
gauge) anomalous effective action is constructed from (vector) gauge
invariant (or covariant) quantities. This is probably the reason why
these two methods lead to an identical result.

We should however emphasize that the results given in
Eqs.\,(\ref{eq59}) and (\ref{eq60}) have a potential ambiguity. This
can be seen from the last limit in Eq.\,(\ref{eq58}), which is, in a
strict mathematical sense, directional dependent, i.e. it may take
distinct values as $x_\mu$ approaches to zero from different
directions. In general, $\lim_{x \rightarrow 0} x_\mu x_\nu/x^2 = C
g_{\mu\nu}$ with $C$ being an arbitrary constant. To ensure the
consistency of the trace in four dimensions, we choose $C=1/4$. It is
in this sense that the value of the radiatively induced Lorentz and
$CPT$ violation is ambiguous in our approach. From a theoretical
point of view, we find the calculations performed using the covariant
derivative expansion and the Schwinger proper-time method more
appealing. Nevertheless, if the radiatively induced Chern-Simons term
has any physical observable effect, it is the comparison with
experiment which will fix such ambiguities and ultimately resolve the
discrepancies among various results.

\acknowledgments

This work is supported by the Academy of Finland under the Project
No. 163394. W.F.C. is also supported by the Natural Sciences and
Engineering Research Council of Canada. W.F.C and R.G.F are grateful
to the Helsinki Institute of Physics, where part of this work was
accomplished, for warm hospitality. We would like to thank Professors
L.H. Chan, V.A. Kosteleck\'{y} and Dr. M. P\'{e}rez-Victoria for enlightening
discussions and communications. We thank R. Jackiw for bringing to
our attention a wrong sign in the first version of the paper and for
discussions.

\end{document}